\documentclass[fleqn,usenatbib]{mnras}

\usepackage{newtxtext,newtxmath}
\usepackage[T1]{fontenc}
\usepackage{ae,aecompl}
\usepackage{multicol}
\usepackage{graphicx}
\usepackage{color}
\usepackage{graphicx}
\usepackage{rotating}
\usepackage{longtable}
\usepackage{lscape}
\usepackage{multirow}
\usepackage{multicol}
\newcommand{\figref}{Figure~\ref}
\newcommand{\tabref}{Table~\ref}

\newcommand{\secref}{Section~\ref}

\title[The disruption of E\,3]{The disruption of the low-mass globular cluster E\,3}

\author[J. A. Carballo-Bello et al.]{Julio A. Carballo-Bello$^{1}$\thanks{E-mail: jcarballo@uta.cl}, Ricardo Salinas$^{2}$ \& Andr\'es E. Piatti$^{3,4}$\\
$^{1}$Instituto de Alta Investigaci\'on, Universidad de Tarapac\'a, Casilla 7D, Arica, Chile\\
$^{2}$Gemini Observatory/NSF's NOIRLab, Casilla 603, La Serena, Chile\\
$^{3}$Instituto Interdisciplinario de Ciencias B\'asicas (ICB), CONICET-UNCUYO, Padre J. Contreras 1300, M5502JMA, Mendoza, Argentina\\
$^{4}$Consejo Nacional de Investigaciones Cient\'ificas y T\'ecnicas (CONICET), Godoy Cruz 2290, C1425FQB, Buenos Aires, Argentina\\
}
\pubyear{2020}

\begin{document}
\label{firstpage}
\pagerange{\pageref{firstpage}--\pageref{lastpage}}
\maketitle

\begin{abstract}
We use {\it Gaia} DR2 photometry and proper motions to search for the hypothetical tidal tails of the Galactic globular cluster E\,3. Using a modified version of a classical decontamination procedure, we are able to identify the presence of an extended structure emerging from the cluster up to $r \sim 1$\,deg from its center, thus suggesting that this poorly studied cluster is undergoing a tidal disruption process. These low surface-brightness structures are aligned with the direction to the Galactic center, as expected for a cluster close to its perigalacticon. Different scenarios to explain the important amount of mass lost by this cluster are discussed.
\end{abstract}


\section{Introduction}

The structure and evolution of Galactic globular clusters (GCs) is affected by the tidal stress exerted by the Milky Way, which varies in time as these systems move along their orbits within the Galaxy, while being exposed to strong interactions with the densest Galactic components  \citep[e.g.][]{Combes1999}. This process led to the formation of stellar streams or tidal tails, such as the ones generated by the disruption of Pal\,5 \citep{Odenkirchen2003,Grillmair2006a,Kuzma2015}; one of the most extended structures observed among the ones generated by the family of Galactic GCs. In recent years, as new datasets have become available, it has been possible to unveil more of these low surface-brightness tails, making evident that this is a common feature in Galactic clusters and that their formation is a manifestation of their orbital parameters and dynamical evolution \citep[see summary of detections and discussion in ][]{Piatti2020}.  

With the arrival of {\it Gaia}, we have a new opportunity to reveal and trace tidal structures across the sky in areas well beyond their tidal radii, by including new parameters (e.g. parallaxes, proper motions) that were not available in previous photometric surveys. Different approaches have been proposed to exploit such a precious dataset with that aim, from a modified version of the classical statistical decontamination procedure \citep{Carballo-Bello2019}, to a 5D mixture modelling technique, which is capable of systematically detecting tidal tails in the surroundings of the most massive halo GCs \citep{Sollima2020}. However, as we move to the low-mass end in the distribution of Galactic GCs, the search for faint tails becomes a difficult task because of the limitations on successfully separating the cluster content from the fore/background stellar populations.
 
 On the other hand, low-mass clusters may \textit{favour} the generation and detection of tidal tails. As shown by \citet{balbinot2018}, the average mass of an escaping star in a low-mass cluster is higher (and therefore brighter) than those in a high-mass cluster, making its tidal tails more clearly visible. This can be the case for low-mass clusters showing hints of formation of tails and/or tidal disruption \citep[e.g. Whiting\,1, Pal\,13 and AM\,4;][]{Carraro2007,Carraro2008,Carballo-Bello2017,Piatti2020b,Shipp2020}.
 
With a present-day mass of only around $3\times10^3 {\rm M}_{\odot}$ \citep[][ see other basic parameters in \tabref{tabla}]{Baumgardt2019}, the star cluster E\,3 \citep[also known as C\,0921-770 and  ESO\,37- 1;][]{Lauberts1976} is one of the least massive GCs in our Galaxy, and belongs to the population of oldest clusters \citep{marin2009}. Unlike the great majority of Galactic GCs, it does not show evidence of multiple stellar populations \citep{Salinas2015,monaco2018}. Since multiple populations are produced by the ability to retain enriched material in the early life of a cluster, an absence of them indicates that the initial mass of the cluster was also lower than the bulk of Galactic GCs. The sparse nature of this cluster, together with a dearth of low mass stars in its color-magnitude diagram (CMD), were noticed early on, hinting at a tidal removal of stars \citep{vdb1980,mcclure1985}, and giving the cluster its nickname of a ``dying globular cluster''. In this work,  we explore {\it Gaia} DR2 data trying to detect the hypothetical tidal tails resulting from the disruption of E\,3. 

\begin{table}
\small
\begin{centering}
\begin{tabular}{ll}
\hline
R.A. &  140.238\,deg\\
Dec &  -77.282\,deg\\ 
$d_{\odot}$ & 8.1\,kpc \\
$d_{\rm GC}$ & 9.4\,kpc \\
$r_{\rm c}$ & 0.8\,arcmin (1.9\,pc) \\
$r_{\rm h}$ & 2.7\,arcmin (6.3\,pc) \\
$r_{\rm t}$ & 10.2\,arcmin (24.1\,pc) \\
Mass & $2.9\times10^3$ M$_{\odot}$\\
$v_{\rm r}$ & 12.6\,km\,s$^{-1}$\\

\hline
\end{tabular}
\caption[]{Basic parameters of the cluster E\,3 \citep{Baumgardt2019}. The radial velocity was taken from \cite{monaco2018}.}
\label{tabla}
\end{centering}
\end{table}

\section{Gaia data}
\label{observations}

The European Space Agency (ESA) mission {\it Gaia} is providing precise positions, kinematics and stellar parameters for more than one billion star, and will help us to understand the origin and evolution of our own Galaxy by exploring its current structure  with unprecedented detail \citep{Gaia2016}. We have used the five-parameter astrometic solution (positions, proper motions and parallaxes) and  ($G,~G_{\rm BP},~G_{\rm RP}$) photometry provided by the second data release of this mission \cite{Gaia2018a} to identify likely members of E\,3 in its surroundings.  

We have retrieved all the information available for an area of the sky within 5\,deg from the center of E\,3. To ensure a good quality photometry and astrometry for all the sources throughout our analysis, we only consider stars with \textsc{phot\_bp\_rp\_excess\_factor} $\leq$  1.5 and \textsc{visibility\_periods\_used} $\geq$ 5.  We also adopted the formalism of the renormalized unit weight error \citep[RUWE;][]{Lindegren2018b} and we assumed that only objects with RUWE $\leq$ 1.4 have an acceptable astrometry. We have used the {\it Gaia} extinction coefficients provided by \cite{Gaia2018a} and the individual E$(B-V)$ values obtained from the \cite{Schlafly2011} maps.

\begin{figure}
  \begin{center}
  \includegraphics[width=\columnwidth]{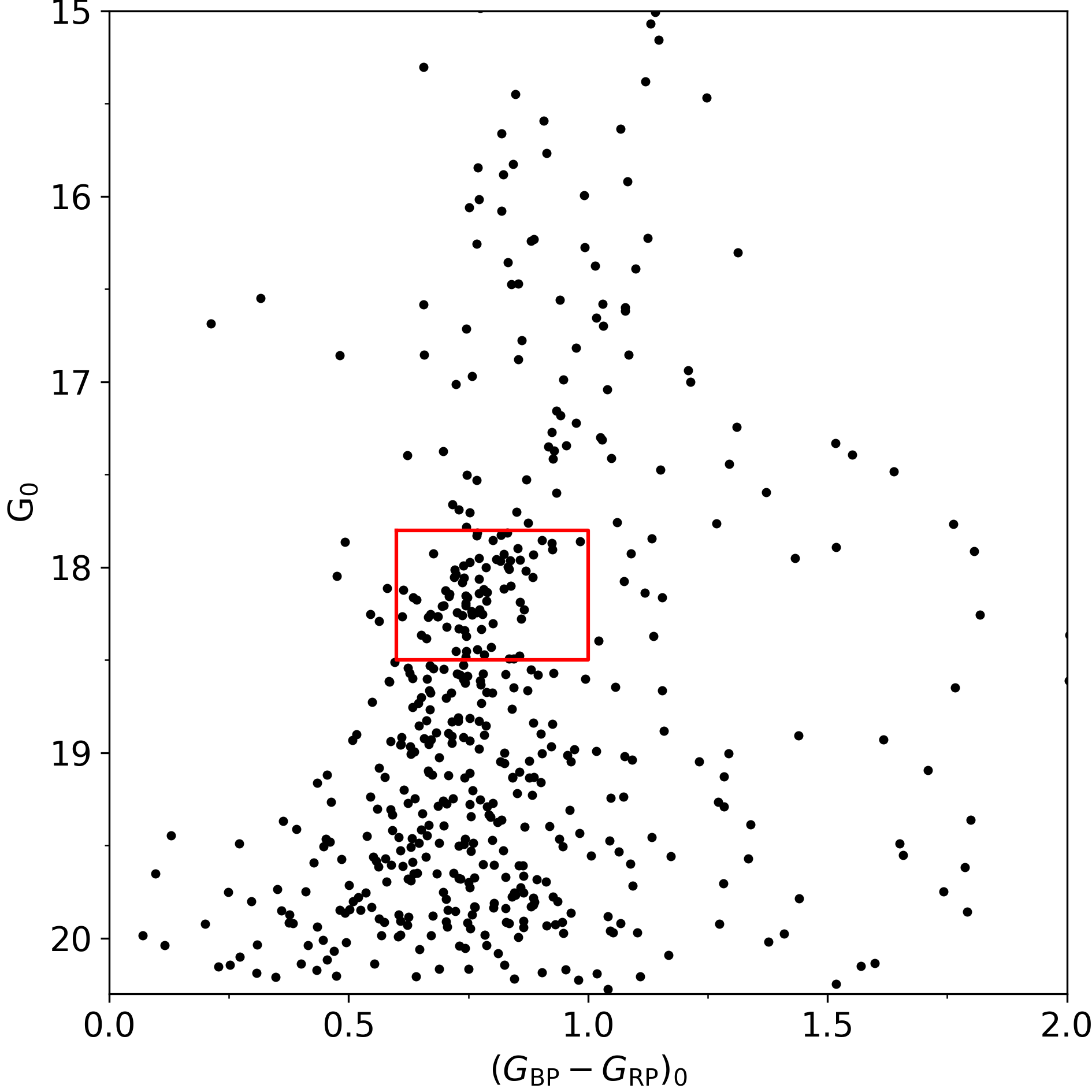}
  \caption[]{CMD corresponding to all the \emph{Gaia} objects within $ r \leq 3$\,arcmin from the center of E\,3 \citep[coordinates from][]{Baumgardt2019}. The red box indicates the position of the objects used to define the $\varpi$, $\mu_{\alpha^{*}}$ and $\mu_{\delta}$ ranges considered in our procedure.} 
 \label{cmd}
    \end{center}
 \end{figure}

\section{Methodology}
\label{methodology}

With the purpose of estimating the probability of each star of belonging to E\,3, we follow the procedure described and used in \cite{Carballo-Bello2019} to unveil extra-tidal features around NGC\,362. This method considers color, magnitudes and proper motions of the stars in the region under study and compare their distribution in the same planes for a sample of control field stars. We select an area around E\,3 of $1.5$\,deg $\times 1.5$\,deg where we expect to identify the tentative members left by the clusters in its surroundings. As control sample, we used those located in a region beyond $r = 1.5$\,deg and with an equivalent total area. The ($G_{0},G_{BP_{\rm 0}}-G_{RP_{\rm 0}},\mu_{\alpha^{*}},\mu_{\delta}$) space is divided into a grid of cells, where the cell size $\epsilon$ are $\epsilon_{\rm G_{\rm BP_{0}}-G_{\rm RP_{0}}} = 0.2, \epsilon_{\rm G_{0}} = 0.5, \epsilon_{\mu_{\alpha^{*}}} = 0.5$ and $\epsilon_{\mu_{\delta}}= 0.5$.  We then compute the weight ($\tau$) for all the stars placed in those cells by using the expression

\begin{equation}
\tau = 1 - \frac{N_{\rm field}\,A_{\rm cluster}}{N_{\rm cluster}\,A_{\rm field}}
\end{equation}\\

where $N$ and $A$ correspond to the number of stars in a given ($G_{0}, G_{\rm BP_{\rm 0}}-G_{\rm RP_{\rm 0}}, \mu_{\alpha^{*}}, \mu_{\delta}$) cell and the total area for each population (cluster or control field), respectively. Unwanted effects in the results due to the way in which we divide the spaces are avoided as much as possible by shifting the grids in each dimension by 1/3\,$\epsilon$, yielding 81 different configurations and weight values. In this work, we add an additional step and  compare our target sample with 1000 randomly selected subsamples of field stars, occupying 20$\%$ of the initial field area. With the latter process we address the likely variation of star density in the surroundings of our area of interest, specially because of its proximity to the Galactic plane. We finally assign to each star the mean value $\overline{\tau}$ resulting from the 81\,000 iterations. 

As shown in \figref{cmd}, E\,3 has a poorly populated main sequence (MS) of stars with $G_{\rm 0} \geq 18$ and only the section of the diagram around its tentative MS turn-off is well defined \citep[see the much deeper CMD in][]{Marcos2015}. This low density of cluster members clearly affects the morphology of  the CMD and an important fraction of fore/background stars are also identified in its brigher and redder section. In order to reduce the number of polluters in our final sample, we limit our method to the ranges in parallax and proper motions defined by the stars likely associated with E\,3  with $ 17.8 \leq G_{\rm 0} \leq 18.5$ and $0.6 \leq (G_{\rm BP}-G_{\rm RP})_{0} \leq 1$  (see selection box in \figref{cmd}). To define those ranges, we have obtained the error-weighted distributions of $\varpi$, $\mu_{\alpha^{*}}$, $\mu_{\delta}$,  for stars with $r \leq 3$\,arcmin  (similar to the $r_{\rm h}$ of this cluster) and $90  \leq r [{\rm arcmin}] \leq 120$ using a bin size of 0.03, 0.3, 0.3, respectively. Parallax values have been corrected by adding a zero point of 0.04\,mas \citep[e.g.][]{Maiz-Apellaniz2019}.

\begin{figure}
  \begin{center}
 \includegraphics[width=\columnwidth]{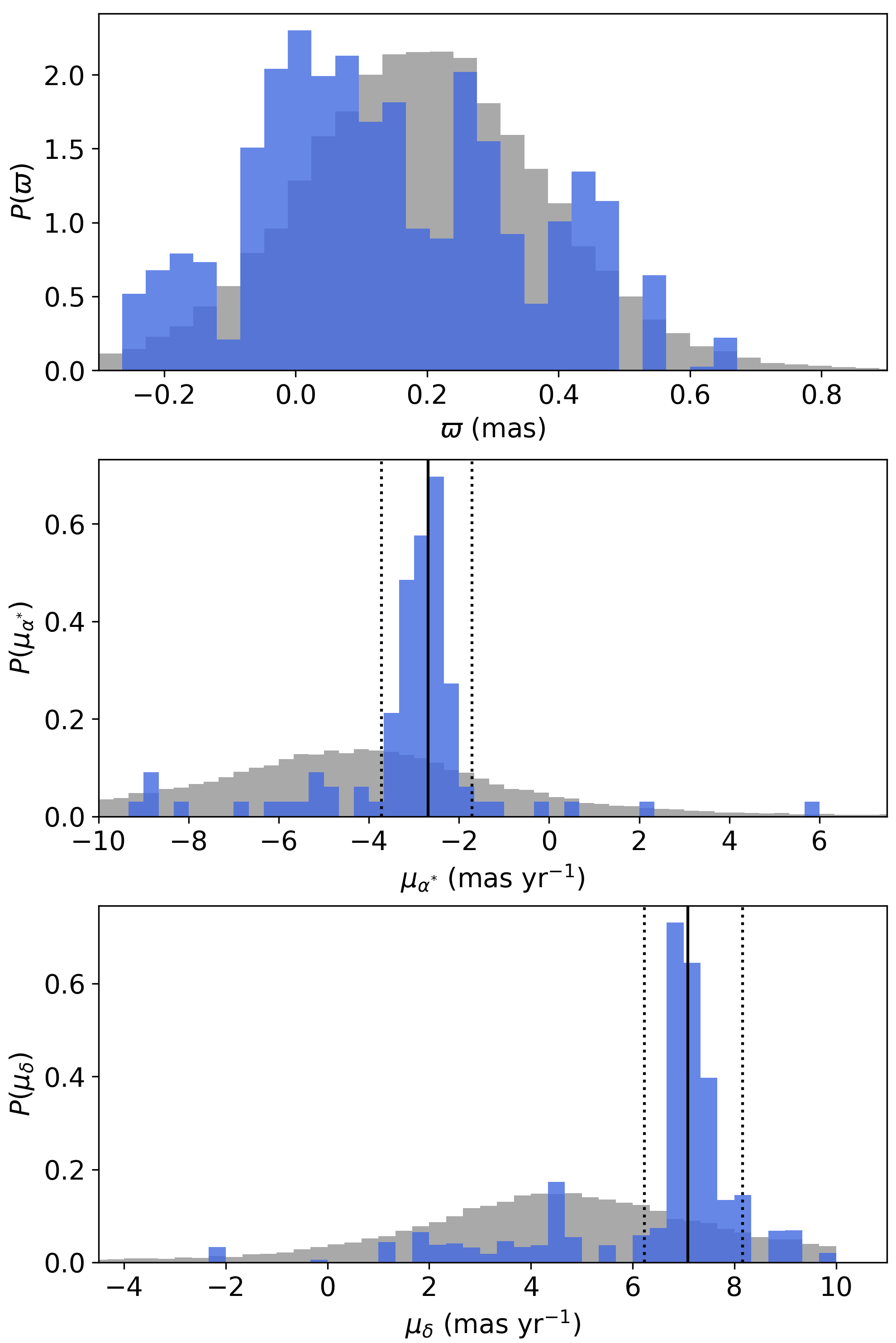}
  \caption[]{Parallax (top) and proper motions (middle and bottom) distributions of stars likely associated with the MS turn-off of the GC E\,3 ( $ r \leq 3$\,arcmin, blue) and field stars ($90 \leq r{\rm [arcmin]} \leq 120$, grey) . The vertical dashed lines indicate the proper motions ranges considered in our analysis, while the solid lines indicate the mean values derived by \cite{Baumgardt2019}.} 
 \label{histograma}
    \end{center}
 \end{figure}

The resulting parallax distribution  (upper panel in \figref{histograma}) shows several peaks and it is not possible to clearly identify the component associated with E\,3. There is a prominent group of stars with values compatible with the mean parallax reported for this GC \citep[$\varpi \sim 0.12$\,mas;][]{Baumgardt2019}. However,  the exclusion of stars with negative or large values may notably bias our sample, thus altering our capacity of detecting the extra-tidal structures \citep[see discussion about usage of \emph{Gaia} DR2 parallaxes in ][]{Luri2018}. In order to avoid the lost of information, specially for such a low-density cluster at $d_{\odot} \sim 8$\,kpc where {\it Gaia} parallaxes have large uncertainties, we do not impose restrictions on the $\varpi$ values. On the other hand, the distributions obtained for $\mu_{\alpha^{*}}$ and $\mu_{\delta}$ and shown in the  middle and bottom panels in \figref{histograma}, respectively, allow us to assume that most of the stars likely associated with E\,3 are contained within 1 standard deviation around the mean values at $\mu_{\alpha^{*}} = -2.7 \pm 0.9$\,mas\,yr$^{-1}$ and $\mu_{\delta} = 7.2 \pm 0.9$\,mas\,yr$^{-1}$, which are in good agreement with the proper motions derived by \cite{Baumgardt2019} for this cluster. Since we expect that most of the stars lost by GC are low-mass MS members, we focus our analysis on the  $0.6 \leq (G_{\rm BP}-G_{\rm RP})_{0} \leq 1$  and $G_{\rm 0} \geq 17.8$ section of the CMD. 

From the initial sample of 161\,733 stars in our area of interest ($r \leq 90$\,arcmin), we proceed in our analysis with a total of 2077 stars satisfying the criteria described above. As for the control field sample, 1463 out of the 127798 objects observed by {\it Gaia} in the $90  \leq r [{\rm arcmin}] \leq 120$ area around E\,3 were used in our method.

\section{Results and discussion}

\begin{figure*}
  \begin{center}
 \includegraphics[scale=0.5]{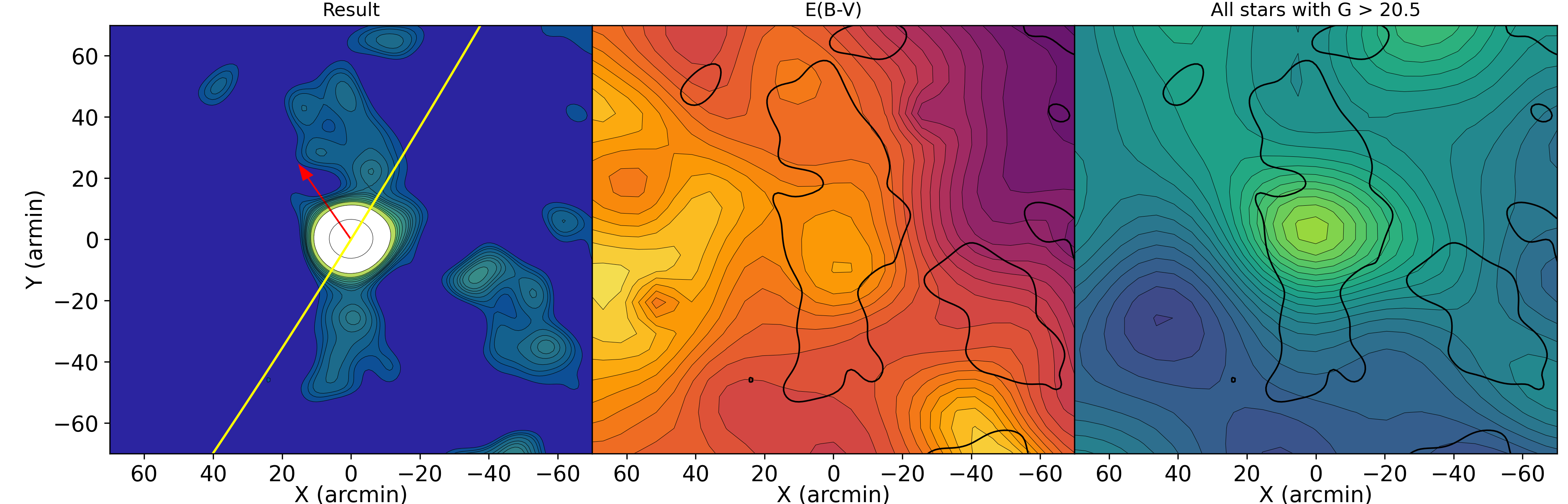}
  \caption[]{{\it Left:} density map generated for the surroundings of E\,3. The plotted contours correspond to the range $1.5 \leq S \leq 5$, with increments of 0.25. The solid yellow line represents the tentative orbit of the cluster, while the red arrow indicates the direction of the Galactic center. {\it Middle:} $E(B-V)$ map for the same sky area, with 30 contour levels in the range $ 0.1 \leq E(B-V) \leq 0.3$. The overplotted black contours correspond to the $S=1.5$ region from the left panel. {\it Right:} distribution of sources originally found in {\it Gaia} DR2 with $ G \geq 20.5$. } 
 \label{mapa}
    \end{center}
 \end{figure*}

The density map shown on the left panel of \figref{mapa} was generated by summing the individual weights in bins of  $5$\,arcmin~$\times~5$\,arcmin. The result was smoothed using a Gaussian filter with a width equivalent to 3 bins and converted into a significance ($S$), which represents the standard deviations over the mean value in the field  ($S = ({\rm signal} - {\rm <signal>})/{\sigma_{\rm signal}}$). At first glance, a low-significance structure is detected in the surroundings of E\,3, which is oriented in the north-south direction and up to distances of $\sim 1$\,deg ($\sim 140$\,pc) from the cluster center  (assuming the \cite{Baumgardt2019} heliocentric distance).  No further limitations have been imposed to $G_{\rm 0}$ in our procedure given that it seems reasonable to expect that fainter stars (with larger photometric errors) will have smaller mean $\tau$ values and a lower impact in the resulting density map. Indeed, the north-south structure unveiled by our technique is observed even when the density map is built with stars with different  $G_{\rm 0 max}$ values in the range $18.5 \leq G_{\rm 0 max} \leq  20$. Moreover, the orientation of these tails is not altered when different bin sizes and/or  filter widths are used. 

While the southern arm, an apparently narrower structure, seems to be better aligned with the cluster center, the northern component seems to be more dispersed, with a lower mean significance, and distributed along an axis which is slightly shifted from the central coordinates of E\,3. Misaligned tails, specially in those sections far away from the cluster center, have been observed in other Galactic GCs exhibiting tidal tails \citep[e.g.][]{Navarrete2017}. We have explored whether these tails are associated with gradients in the extinction or the distribution of {\it Gaia} DR2 sources over the field. As shown in the middle panel in \figref{mapa}, there is a variation in the \cite{Schlafly2011} $E(B-V)$ values, with a maximum and minimum values of $E(B-V) \sim 0.3$ and $0.1$, respectively, with a $\sigma \sim 0.06$. Therefore, although we observe a few extinction peaks in our area of interest, those variations are not reflected in our results. Moreover, since completeness of {\it Gaia} might be affected, among other factors, by its scanning laws,  we also checked for variations in the density of stars throughout our field by counting stars with $G \geq 20.5$ in the original catalog \citep[see analysis of the completeness of {\it Gaia} in][]{Boubert2020}. Besides the expected smooth gradient of star counts towards higher Galactic latitudes and the Milky Way plane (see right panel in \figref{mapa}), there are no hints of incompleteness in the faint end of the objects observed by {\it Gaia}. We thus conclude this structure seems to be a real overdensity of stars likely associated with E\,3.

\begin{figure}
  \begin{center}
  \includegraphics[width=\columnwidth]{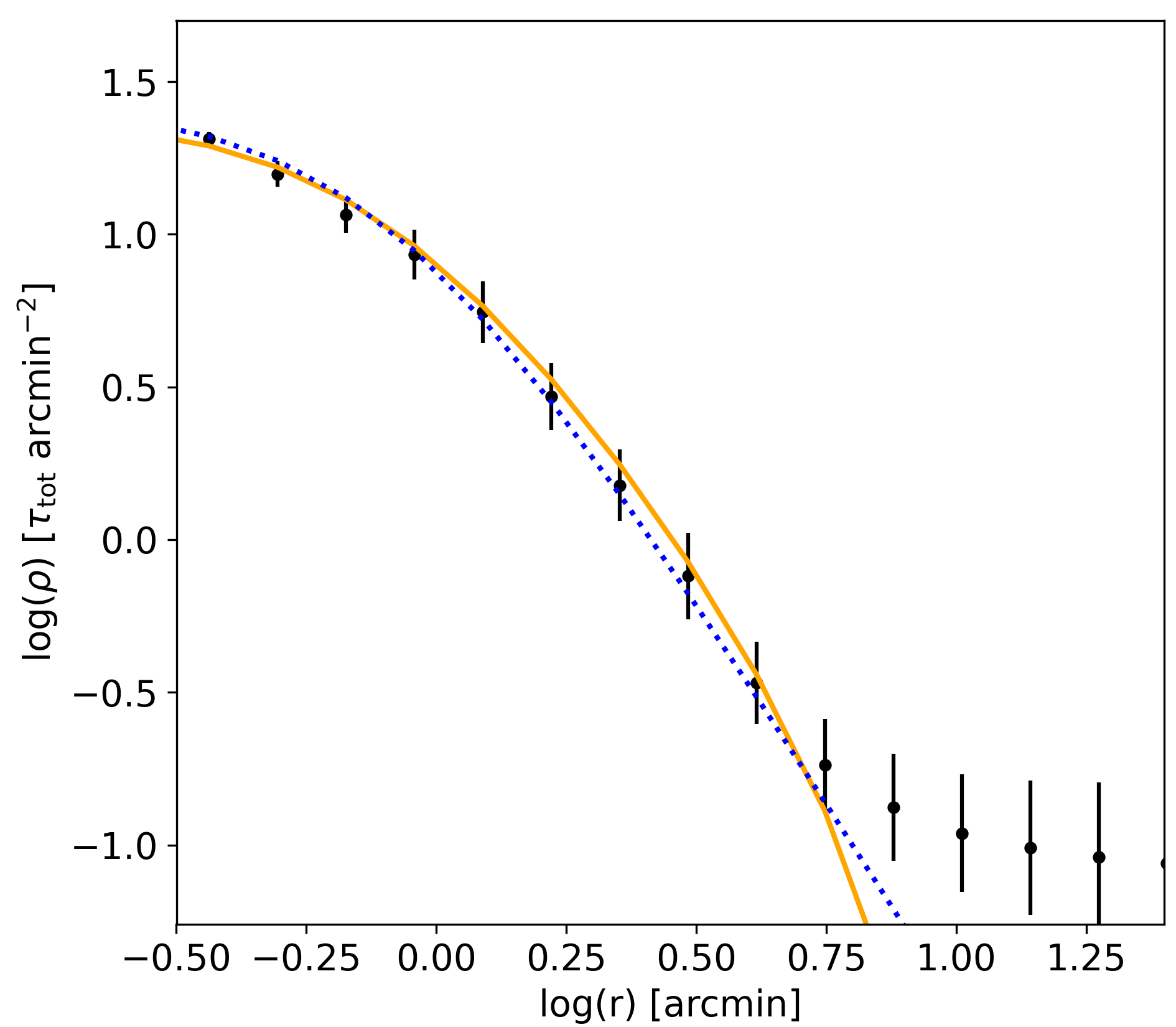}
  \caption[]{Radial density profile of E\,3 generated from the sum of the individual weights. The solid yellow and dashed blue lines correspond to the \cite{King1966} and \cite{Elson1987} templates fitting, respectively.} 
 \label{profile}
    \end{center}
 \end{figure}

The dynamical evolution and the interaction of the Galactic GCs with a varying tidal field along their orbits around the Milky Way reflects in their overall structure. Indeed, a population of potential escapers is built in the outer regions of the clusters during their evolution \citep[see references and discussion in][]{deBoer2019}, which manifest in the observation of a break or change in the slope of the density profile. This become more relevant in those clusters with emerging tidal tails, where the surface density profile slope is shallower than for the bulk of GC members \citep[e.g. Pal\,5 and NGC\,5466, ][respectively]{Odenkirchen2003,Belokurov2006}. Since we only have individual weights for all the stars in the field instead of a decontaminated catalog, we generated the radial distribution shown in \figref{profile} by summing their weights over concentric rings centered in E\,3 rather than counting cluster stars. In order to locate the tentative position of the break in the profile, which may indicate the presence of potential escapers or/and the presence of tidal tails as the one observed in \figref{mapa}, we fitted \cite{King1966} and \cite{Elson1987} models to the distribution within $r = 8$\,arcmin. The bulk of stars ($ r \sim 5$\,arcmin) is well fitted by both models with $r_{\rm c} = 0.7 \pm 0.1$\,arcmin  and $r_{\rm t} = 12 \pm 0.2$\,arcmin for the King model and with $r_{\rm eff} = 0.8 \pm 0.1$\,arcmin and $\gamma = 2.7 \pm 0.2$ for the Elson power-law template. The structural parameters derived here are in good agreement with those obtained by \cite{Baumgardt2019}.  Beyond $r \sim 5$\,arcmin, the deviation of the observational profile from both fits suggests the existence of a group of potential escapers, followed by a distribution that slowly decays up to distances much larger than the King tidal radius of E\,3.

There is enough evidence showing that the morphology of the tidal tails in a GC is affected by the orbit followed by the cluster \citep[e.g.][]{Montuori2007,Carballo-Bello2012,Kupper2012,Piatti2020,Sollima2020}. More specifically, numerical simulations and the systematic search for extratidal content in Galactic GCs have shown they are often aligned with the orbit of the cluster, specially in the regions well beyond its tidal radius \citep[e.g.][]{Montuori2007,Klimentowski2009}.  In order to confirm whether it is also the case of these tails, we compute a tentative orbit for E\,3 using the \textsc{galpy} package \citep{Bovy2014} and assuming the \cite{Reid2014} values for the distance of the Sun to the  Galactic center and its circular velocity set at $R_{\odot} = 8$\,kpc and $V_{\odot} = 240$\,km\,s$^{-1}$, respectively. As for the cluster, we have used the mean proper motions derived from \figref{histograma}, the radial velocity from \cite{monaco2018} and set at $v_{\rm r} = 12.6$\,km\,s$^{-1}$ , and the heliocentric distance estimated by \cite{Baumgardt2019}.

The resulting orbits are overplotted on the left density map shown in \figref{mapa}. The tidal tails unveiled in this work are found well beyond its tidal radius ($r_{\rm t} = 24$\,pc), poorly correlated with the orbit of the cluster but aligned with the direction to the Galactic center. E\,3 has crossed its perigalacticon at $d_{\rm GC per} \sim 9$\,kpc only $\sim 20$\,Myr ago, thus  according to \cite{Montuori2007},  the inner tidal tails should point towards the Galactic center and they are only good tracers of the orbital path at large scales.  Since we are not able to reveal any other overdensities in the surroundings of E\,3 following the methodology described in \secref{methodology}, even when we increase the area analysed, we may suggest that the orientation of the structures detected here results more from the orbital stage of the cluster than from a reflection of its path around the Milky Way. It is also important to emphasize the difficulties found to properly determine the main basic parameters associated with the orbit of such a faint GC, as evidenced by the difference between the reported radial velocities measurements for this cluster \citep{Marcos2015,Salinas2015,monaco2018}. 

Despite the low eccentricity ($e = 0.2$) of the orbit in which E\,3 is placed, this cluster has lost a remarkable fraction of stars  compared to the rest of Galactic GCs with similar orbital parameters \citep{Piatti2020}. Its core radius is also comparable or larger than the ones observed in the other members of that same family of GCs, and dynamically looks like a more evolved cluster. Using Eq. 5 in \cite{Piatti2019masa}, we estimate that the  fraction of cluster mass lost by tidal heating for E\,3 is $M_{\rm dis} / M_{\rm ini} = 0.42$. Assuming that all the stars in our field have similar masses and that the extended low-density tails start at $r \sim  15$\,arcmin (based on a visual inspection of the left panel in \figref{mapa}), we sum weights in the area enclosed by the $S = 1.5$ contour and estimate that the outer structures contain, as an upper limit, a total mass equivalent to the $\sim 40\%$ of the current mass of E\,3 (around a half of the stars lost by the cluster). Such an important mass loss could have been thus originated in a more complex encounter between this GC and any of the Milky Way components, when most of the low-mass stars were ripped out and only the core of original stellar system survived.

In this context, E\,3 may have been formed within an already accreted dwarf galaxy. The Galactic halo, which is mostly result of the continuous merging and accretion of minor 
 satellites \citep[e.g.][]{Rodriguez-Gomez2016}, is populated by a progeny of tidal streams and overdensities.  The Helmi streams, a family of stellar substructures in the Solar neighbourhood \citep{Helmi},  represents an important source of stars in the Galactic halo ($\sim 15\%$) and may also contributed with GCs, which are now members of the Milky Way GC system \citep{Koppelman2019}. Although E\,3 was not included in the initial sample of GCs likely accreted by the Milky Way, \cite{Massari2019}  added this cluster in the candidates list based on its orbital properties and a less restrictive selection criterion. Therefore,  the violent process that partially dissolved E\,3 may correspond to the accretion of a massive galaxy ($M_{\*} \sim 10^{8} M_{\odot}$) into the Milky Way, whose disruption led to the formation of the Helmi streams. Interestingly, the orbit of this cluster crosses the paths in the sky of the Eastern Banded Structure, the Anticenter Stream \citep{Grillmair2006d}, and the Monoceros ring \citep[e.g.][]{Slater2014}, which seem to result from the distortion of the Galactic disk due to its interaction with satellite stellar systems \citep{Deason2018,Laporte2020}, including the Sagittarius dwarf galaxy. Favouring its extra-Galactic origin, accreted GCs without a clear surviving progenitor galaxy as in the case of E\,3 are often found surrounded by extended stellar structures \citep[e.g.][]{Carballo-Bello2018}. 

The detection of tidal tails in low-mass systems such as E\,3 allows us to gain insights into the  the disruption and survival of GCs in the Milky Way, and how their evolution is probably  related to their extra-Galactic origin. Future {\it Gaia} data releases  will provide an opportunity to locate the hypothetical stellar stream originated by the violent disruption of this cluster.

\section{Conclusions}
E\,3 represents a unique case of a Galactic GC on a orbit with low inclination and eccentricity, with an important mass loss  due to a single (or several) episodes during its evolution. In this work, we have tried to unveil the hypothetical tidal tails around this cluster by applying a very restrictive version of a procedure designed to identify likely members of Galactic GCs beyond their tidal radii. 

Our results show that a low-significance substructure emerging from the cluster is aligned with the direction towards the Galactic center, as expected for clusters which are close to their perigalacticon. However, that substructure doest not contain enough stars to account for the mass lost by E\,3. Future {\it Gaia} data releases might allow us to trace the rest of the tidal structure generated by the disruption of this cluster and establish whether the survival of this GC is related to its evolution within an accreted dwarf galaxy or peculiar born conditions. 

\section*{Data Availability}

This work has made use of data from the European Space Agency (ESA) mission {\it Gaia} (\url{https://www.cosmos.esa.int/gaia}), processed by the {\it Gaia} Data Processing and Analysis Consortium (DPAC, \url{https://www.cosmos.esa.int/web/gaia/dpac/consortium}). Funding for the DPC has been provided by national institutions, in particular the institutions participating in the {\it Gaia} Multilateral Agreement.

\section*{Acknowledgements}

Thanks to the anonymous referee for her/his helpful comments and suggestions.

\def\jnl@style{\it}                       % Defines journal style in italics
\def\mnref@jnl#1{{\jnl@style#1}}          % Defines \mnref command to call journals
\def\aj{\mnref@jnl{AJ}}                   % Astronomical Journal
\def\apj{\mnref@jnl{ApJ}}                 % Astrophysical Journal
\def\aap{\mnref@jnl{A\&A}}                % Astronomy and Astrophysics
\def\apjl{\mnref@jnl{ApJL}}               % Astrophysical Journal, Letters
\def\mnras{\mnref@jnl{MNRAS}}             % Monthly Notices of the RAS
\def\nat{\mnref@jnl{Nat.}}                % Nature
\def\iaucirc{\mnref@jnl{IAU~Circ.}}       % IAU Circulars
\def\atel{\mnref@jnl{ATel}}               % Astronomers Telegram
\def\iausymp{\mnref@jnl{IAU~Symp.}}       % IAU Symposium
\def\pasp{\mnref@jnl{PASP}}               % Astronomical Journal
\def\araa{\mnref@jnl{ARA\&A}}             % Astronomical Journal
\def\apjs{\mnref@jnl{ApJS}}               % Astronomical Journal
\def\aapr{\mnref@jnl{A\&A Rev.}}          % Astronomical Journal

\bibliographystyle{mn2e}
\bibliography{biblio}

% Don't change these lines
\bsp	% typesetting comment
\label{lastpage}
\end{document}